\documentstyle[twoside,fleqn,espcrc2,epsf]{article}

\def\mv{m_{\rm val}}
\def\ms{m_{\rm sea}}
\def\pbp{\langle\overline\psi\psi\rangle}
\def\pqpbp{\langle\overline\psi\psi(\mv)\rangle_{\beta,\ms}}

\def\Du{D_{[U]}}
\def\ra{\rangle}
\def\la{\langle}
\def\tr{{\rm Tr}}
\def\Nt{{\rm N_t}}
\def\m{ m }
\def\Sg{{\rm S_{g}}}
\def\Sf{{\rm S_{f}}}

\def\Om{\Omega}

\def\lm{\lambda}
\def\rhulm{\rho_{[U]}(\lm)}
\def\rhob{{\overline\rho}}

\def\beq{\begin{equation}}
\def\eeq{\end{equation}}

\newcommand{\AmS}{{\protect\the\textfont2
  A\kern-.1667em\lower.5ex\hbox{M}\kern-.125emS}}

\title{ Critical behavior of the chiral condensate \\
	at the QCD phase transition
\thanks{This work was supported in part by the US department of
	energy.}
	}

\author{Shailesh Chandrasekharan\address{Department of Physics, 
        Columbia University, New York, NY 10027, USA}%
        \thanks{This work was done in collaboration with D. Chen, N.
		Christ, R. Mawhinney, W. Lee, and D. Zhu.}}
       
\begin{document}

\def\thepage{CU--TP--663 \ \ \  hep-lat/9412070}
\thispagestyle{myheadings}

\begin{abstract}
We study the critical behavior of the chiral condensate near the QCD
phase transition in the background of two fixed light dynamical (sea) quarks. 
We study the condensate for \hfill $5.245 \leq \beta \leq 5.3$ \hfill and 
\hfill $10^{-10}\leq \mv \leq 10$ \\ (in lattice units) on a 
$16^3\; \times \; 4$ lattice using staggered fermions with $\ms$ fixed at 
$0.01$.
\end{abstract}

\maketitle

\section{INTRODUCTION}

  Simulations of the QCD chiral phase transition with dynamical quarks have 
been carried out in the past for a variety of quark masses at different
lattice spacings\cite{qcd_ph_rev}. For two light flavors of staggered fermions
these simulations indicate a smooth transition. If this turns out to be a
second order transition, one expects the associated critical behavior to be 
in the $O(4)$ universality class\cite{rajagopal}. There have been 
attempts\cite{karsch} to fit the 
existing data with the $O(4)$ exponents with some success. However to gain 
confidence in such fits one must understand the physics better, especially in 
the range of parameters used in these simulations. For example, when
the quark masses are sufficiently large, a weak first order signal may not 
be resolved and could be seen as a smooth transition. In order to answer this 
question and understand 
the physics of the transition from a slightly different angle we have asked a 
somewhat new question; how does the transition look when studied in the small 
$\mv$ (valence-quark mass) limit, for a small but fixed $\ms$ (sea-quark mass)?

\section{CHIRAL CONDENSATE}

  The order parameter of QCD is the chiral condensate $\pbp$ which is given 
by the formula 
\beq
\pbp(\m,\beta) \;=\; {1\over \Om}\;
\left<\tr{1\over \Du + \m}\right>_{\beta,\m}
\eeq
Here $D_{[U]}$ is the Dirac operator on the lattice for the gauge 
configuration $[U]$ and $\m$ the quark mass in lattice units. The 
expectation value is 
taken over the distribution of gauge configurations given by the usual QCD
path integral
\beq
\la {\cal O} \ra_{\beta,\m} \;=\;
\int d[U] \; {\cal O}[U]\; 
\exp\left[ -\beta\;\Sg+ {\Sf}(\m)\right]
\eeq
where $\Sg$ is the Wilson gauge action with $\beta=6/g^2$ and 
$\exp[{\Sf}(\m)]$ is the staggered fermion determinant for two quark 
flavors of mass $\m$.
In our convention, $\pbp$ is normalized to $1/\m$ for large $\m$ by 
choosing $\Om$. 

\subsection{Partial quenching}
Distinguishing the masses that enter (1) implicitly through the fermion 
determinant and explicitly through the trace of the propagator, one gets 
\beq
\pqpbp = {1\over \Om}
\left<\tr{1\over \Du + \mv}\right>_{\beta,\ms}
\eeq
In this study we compute $\pqpbp$ for various values of $\mv$ at a small but 
fixed value of $\ms$ near the critical coupling $\beta_c$. This is relatively 
easy in 
comparison to changing the values of $\mv$ and $\ms$ simultaneously as 
required in full 
QCD.  However such studies are familiar at zero temperature (far below the 
critical coupling) and is referred to as the partially-quenched 
approximation.  In this work we have extended such a study to finite 
temperatures especially to get a better understanding of the phase transition.
$\pqpbp$ can be referred to as the partially-quenched chiral condensate. 

\subsection{Dirac spectral density}
If one defines $\rhulm$ as the
density(per unit volume) of eigenvalues of the anti-hermitian Dirac 
operator $\Du$ at the eigenvalue $i\lm$, then one can rewrite (3) as
\beq
\pqpbp \;=\; \int_{-\infty}^{\infty}\; {\rm d}\lm\; 
	{\mv\;\rhob_{\beta,\ms}(\lm)\over \lm^2 + \mv^2}
\eeq
where $\rhob_{\beta,\ms}(\lm) \propto \langle \rhulm \rangle_{\beta,\ms}$.
In the $\mv=\ms \rightarrow 0$ limit we obtain the familiar Banks Casher 
formula$\cite{banks}$
\beq
\lim_{\mv=\ms\rightarrow 0}\;
{\pqpbp} \;=\; \pi\;\rhob_{\beta,0}(0)
\eeq
When $\rhob_{\beta,0}(0) \neq 0$ we have chiral symmetry breaking. Furthermore
$\rhob_{\beta,\ms}(0)$ is expected to be a regular function of $\ms$ for
small $\ms$ at least far from the critical coupling where chiral symmetry 
restoration takes place. Hence we expect $\rhob_{\beta,\ms}(0) \neq 0$ below
the critical coupling and zero above. This suggests that 
$\rhob_{\beta,\ms}(\lm)$ will show a striking critical behavior around the 
critical coupling $\beta_c$. This critical behavior can be observed in 
$\pqpbp$ due to (4).
%
\begin{figure}[htb]
\epsfxsize=75mm
\epsfysize=75mm
\hspace{-3mm}
\epsffile{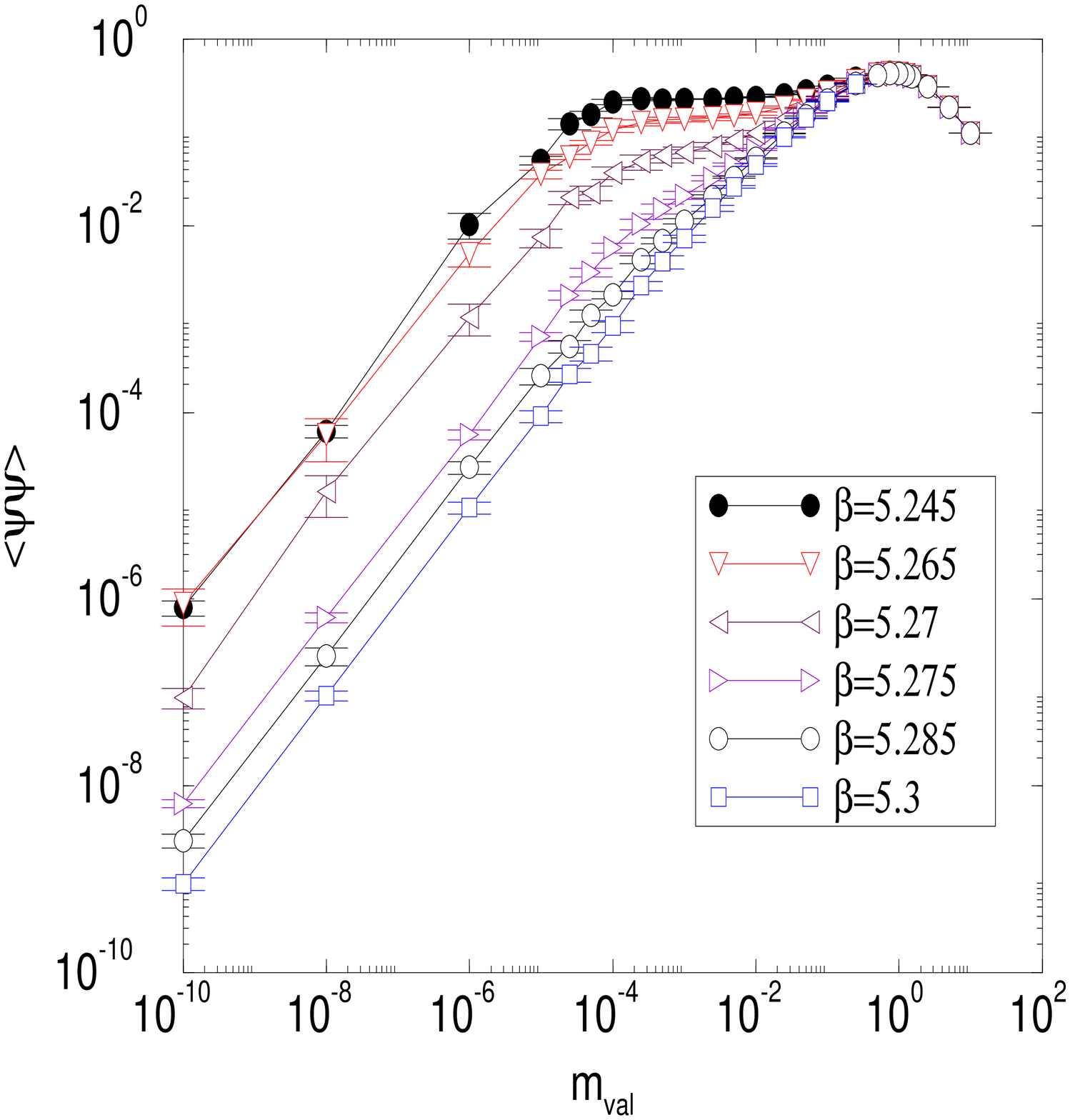}
\vspace{-10mm}
\caption{A log-log plot of $\pbp\; \hbox{vs.}\;\mv$ for different values
of $\beta$ at a fixed $\ms = 0.01$. The figure clearly shows finite volume 
and finite lattice spacing effects. }
\label{fig:loglog}
\end{figure}
  \pagenumbering{arabic}
  \addtocounter{page}{1}
\section{CRITICAL BEHAVIOR}

  In \hfill this \hfill study \hfill we \hfill have \hfill computed  \\ 
$\pqpbp$ using (3) at a fixed 
$\ms = 0.01$ in a spatial volume of $16^3$ and $\Nt = 4$. We have studied 
$\pqpbp$ for $10^{-10} \leq \mv \leq 10$ which
allows us to easily identify the finite volume and finite lattice spacing 
regions. We have simulated at nine different couplings in the region
$5.245 \leq \beta \leq 5.3$. Figure \ref{fig:loglog} shows some of these 
results on a log-log plot.

\subsection{Finite lattice effects}
\begin{figure}[htb]
\epsfxsize=75mm
\epsfysize=75mm
\hspace{-3mm}
\epsffile{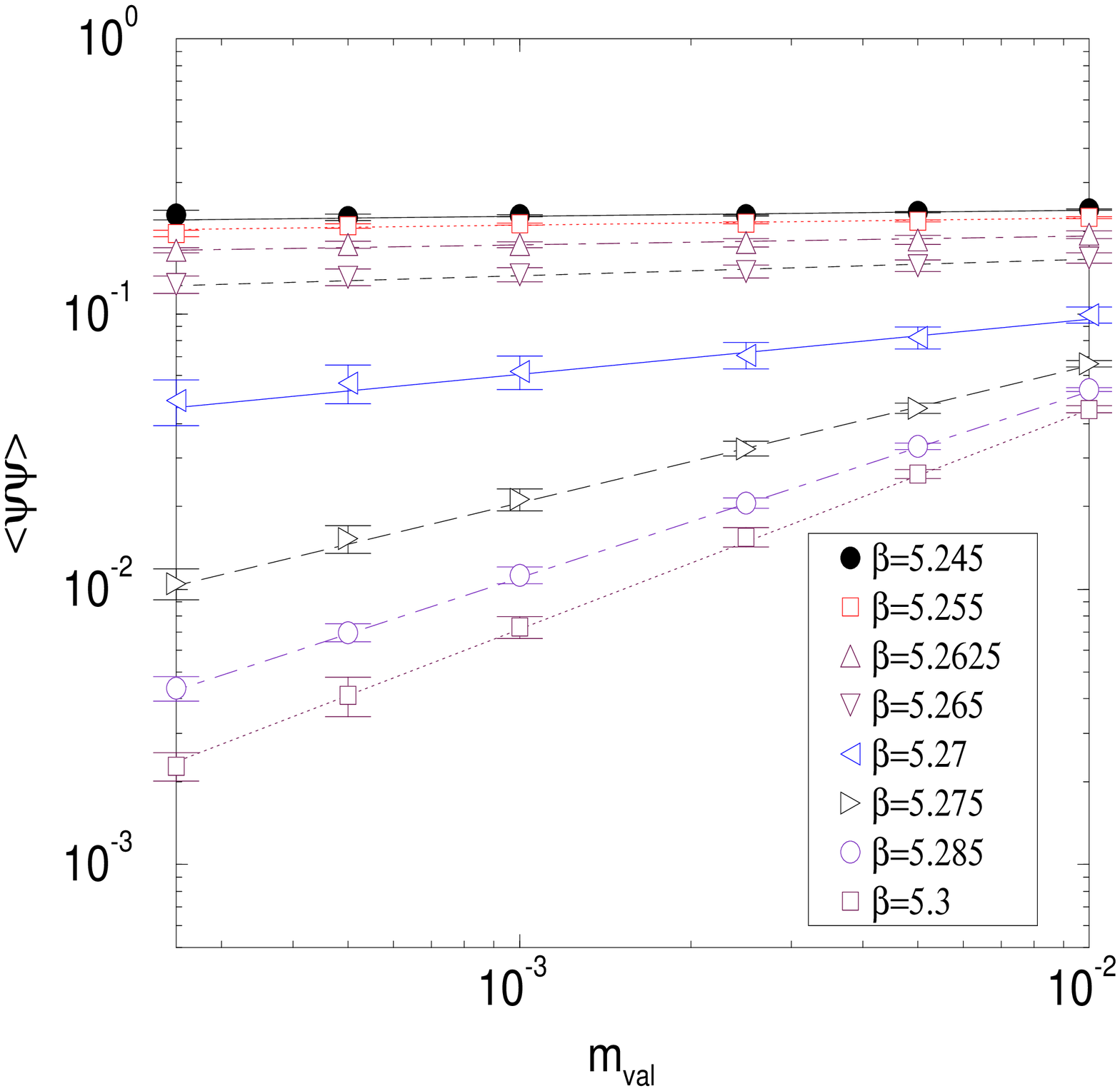}
\vspace{-10mm}
\caption{ The plot of $\pbp\;\hbox{vs.}\;\mv$ is consistent with power-law 
behavior over a range of couplings, in the region free of finite lattice 
effects.}
\label{fig:critical_region}
\end{figure}

The range of $\mv$ used in our study gives us an idea of the
finite lattice-spacing and finite volume effects. We see that for 
$\mv \geq 0.01$ finite lattice spacing effects become important and
for $\mv \leq 2.5\times 10^{-4}$ finite volume effects begin to
set it. One knows that in a finite volume for sufficiently small masses
the condensate must go linearly to zero. This is convincingly seen in
Figure \ref{fig:loglog} for sufficiently small masses. The region 
$2.5\times 10^{-4} \leq \mv \leq 10^{-2}$ could be taken to be
free of the finite size and finite lattice spacing effects. This region is 
plotted for a range of couplings around the critical coupling on Figure
\ref{fig:critical_region}. 

\vspace{-10mm}

\subsection{Power-law behavior}
\begin{figure}[htb]
\epsfxsize=75mm
\epsfysize=75mm
\hspace{-3mm}
\epsffile{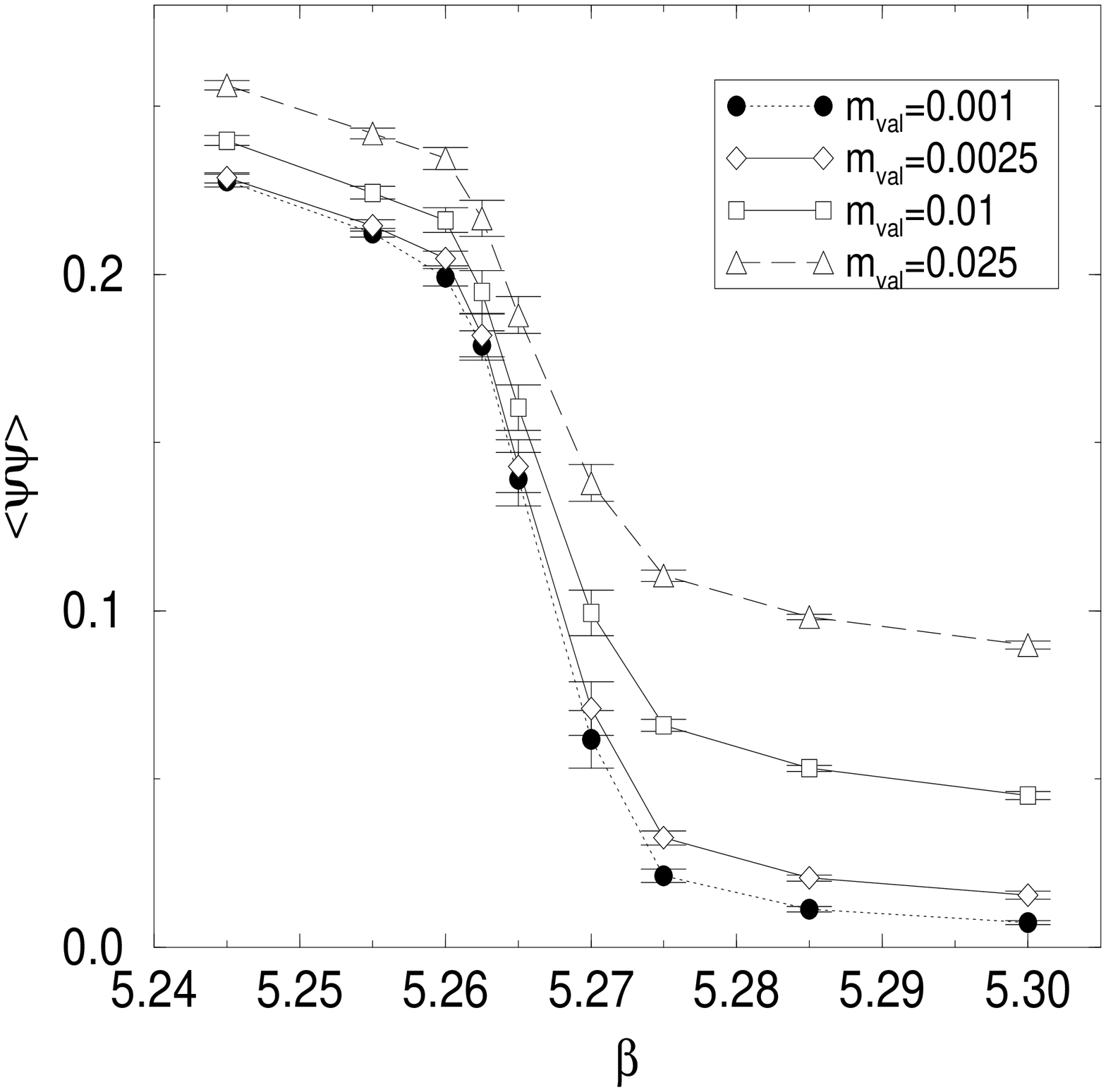}
\vspace{-13mm}
\caption{The plot of $\pbp\;\hbox{vs.}\;\beta$ for various values $\mv$ at a 
fixed $\ms = 0.01$ shows a smooth transition.}
\label{fig:psi_vs_beta}
\end{figure}

   In a typical second order transition one expects a pure power law of the 
form $\pbp \sim m^{1\over \delta}$ at the critical point. It is interesting
to see if such a form also describes the dependence of $\pqpbp$ as 
$\mv \rightarrow 0$. With the available statistics we find that using a 
simple least squares fit the data 
is consistent with a power law over a range of couplings with the power 
${1\over \delta}$ ranging from about 0.02 at a $\beta$ of 5.245 to 0.8 
at a $\beta$ of 5.3. These fits are shown on Figure \ref{fig:critical_region}. 
The plot of $\pbp\;\hbox{vs.}\;\beta$ for a range of $\mv$ is given in Figure 
\ref{fig:psi_vs_beta}.
Here we again see evidence for a smooth transition with the critical 
coupling between 5.26 and 5.275. 
\vspace{-10mm}
\section{CONCLUSIONS}
\begin{figure}[htb]
\vspace{-10mm}
\epsfxsize=75mm
\epsfysize=75mm
\hspace{-3mm}
\epsffile{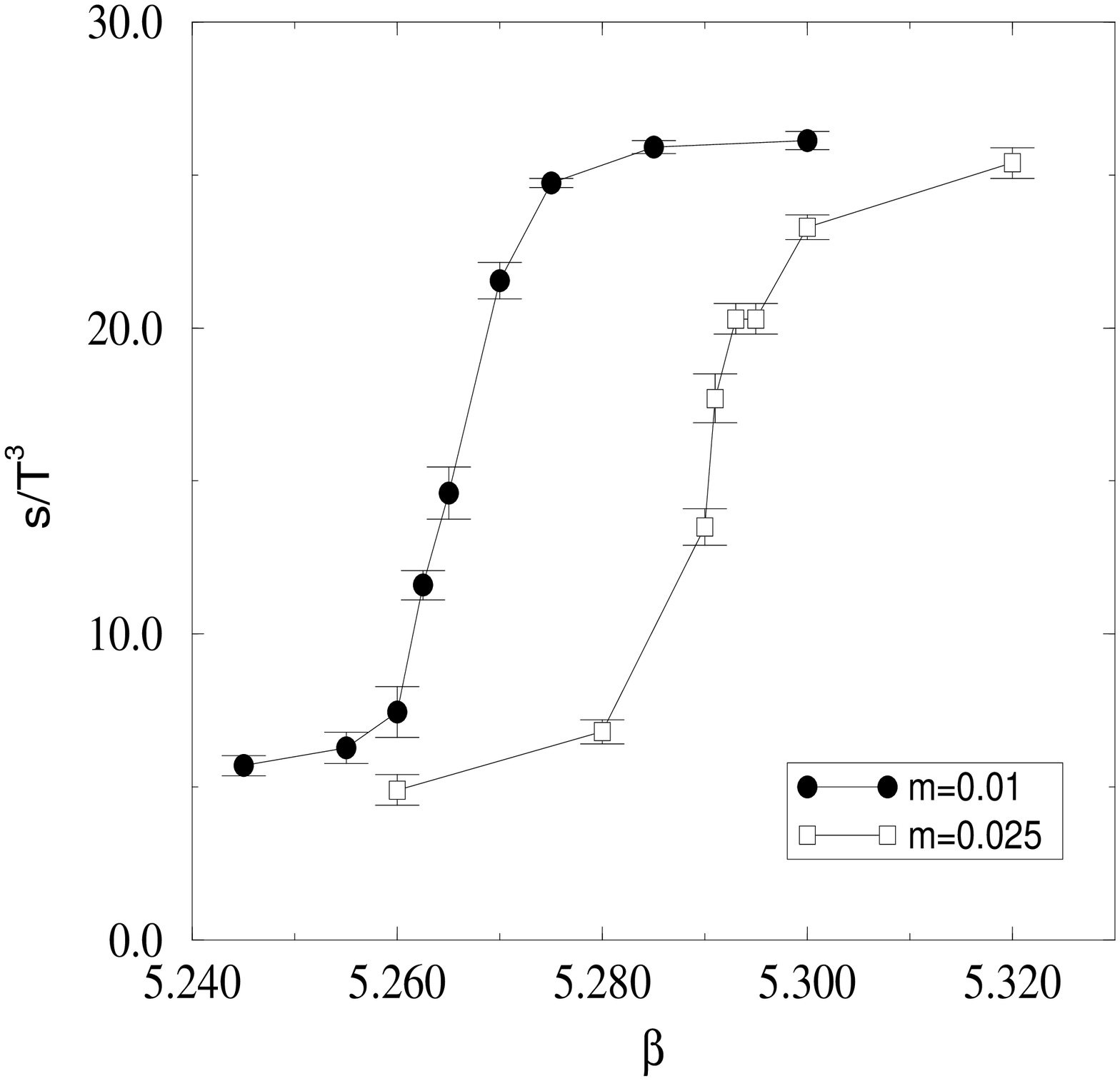}
\vspace{-13mm}
\caption{The plot of entropy density $\hbox{vs.}\;\beta$ at $\ms=\mv$ of 0.01
and 0.025.}
\label{fig:entropy}
\end{figure}

  The data suggests that at  $\ms=0.01$ there is no
discontinuity in the distribution of gauge configurations as a function
the coupling. Any possible discontinuity will be enhanced by our method of 
measuring the condensate at very small valence quark masses. Hence this study 
supports our earlier results that at $\ms=0.01$ there is no 
evidence of a first order transition. Given the accurate results that we have
obtained for $\ms = 0.01$ close to the critical region we can compare with our
earlier results for $\ms = 0.025$. Figure \ref{fig:entropy} is a plot  of 
the entropy density from the present run and the earlier run at 
$\ms$ of 0.025. 
Except for a change in $\beta_c$ caused by the variation of $\ms$, the
sharpness in the two graphs appear remarkably similar unlike the graphs for 
$\mv = 0.01$ and $0.025$ in Figure \ref{fig:psi_vs_beta} where $\ms = 0.01$ is 
fixed. This comparison of Figure \ref{fig:psi_vs_beta} and Figure 
\ref{fig:entropy} also shows that the sea-quark effects are 
the dominant effects in the analysis of pseudo-critical coupling done by 
Karsch\cite{karsch}. Studying the effects of $\ms$ and $\mv$ separately 
will be helpful in understanding some of these observations and can make 
partially quenched studies a useful technique in learning about the real 
critical region.

\end{document}